\documentclass[aip,pop,superscriptaddress,amsmath, amssymb,reprint]{revtex4-1}
\usepackage{graphics}
\usepackage[sort&compress]{natbib}
\usepackage{epsfig}
{\bibpunct{}{}{,}{s}{}{}}

\usepackage[pdftex,breaklinks=true,bookmarksopen=true,bookmarksopenlevel=3,bookmarksnumbered=true,colorlinks=true,urlcolor= magenta,citecolor=blue,linkcolor=blue]{hyperref}

\begin{document}
\title{Tuning the electron energy by controlling the density perturbation position in laser plasma accelerators}

\begin{abstract}

A density perturbation produced in an underdense plasma was used to improve the quality of electron bunches
produced in the laser-plasma wakefield acceleration scheme. Quasi-monoenergetic electrons were generated by
controlled injection in the longitudinal density gradients of the density perturbation. By tuning the position
of the density perturbation along the laser propagation axis, a fine control of the electron energy from a 
mean value of $60$ MeV to $120$ MeV has been demonstrated  with a relative energy-spread of $15 \pm 3.6\%$, 
divergence of $4 \pm 0.8$ mrad and charge of $6 \pm 1.8$ pC.

\end{abstract}
\author{P. Brijesh}
\author{C. Thaury}
\author{K. Ta Phuoc}
\author{S. Corde}
\author{G. Lambert}
\author{V. Malka}
\affiliation{Laboratoire d'Optique Appliqu\'ee, ENSTA ParisTech - CNRS UMR7639 - Ecole Polytechnique, 91761 Palaiseau,
France}
\author{S.P.D. Mangles}
\author{M. Bloom}
\author{S. Kneip}
\affiliation{Blackett Laboratory, Imperial College, London SW7 2AZ, United Kingdom}
\maketitle

\section{INTRODUCTION}
%\vspace{-0.5cm}

\paragraph*{} Higher energy gains, reduced energy spread, smaller emittance and better stability of laser-
plasma accelerated electrons\cite{Tajima1979PRL} are critical issues to address for future development and
applications of compact particle accelerators and radiation sources\cite{Malka2008Nature}. Plasma density
perturbation as a means of controlling electron acceleration\cite{Takada1984AppPhyLett, Bulanov1993LaserPhy} 
and electron injection\cite{Bulanov1998PRE} in laser-generated plasma wakefield is an active research topic. 
Radial density gradients can modify the structure of wakefields and influence the process of injection by 
wavebreaking through a dependence of the plasma wavelength on the transverse co-ordinate\cite{Bulanov1997PRL}. 
Injection of electrons into a narrow phase space region of the wakefield is necessary for generating 
accelerated electron bunches with good quality in terms of energy spread and divergence. Decreasing density 
profile along the laser propagation direction can lead to controlled injection of electrons and is expected to 
generate electron bunches with better beam quality in comparison to self-injection in a homogeneous plasma by 
reducing the threshold of injection within a narrow phase region of the 
wakefield\cite{Bulanov1998PRE,Fubiani2006PRE}. \nolinebreak[4]

\paragraph*{} Injection in a longitudinally inhomogeneous plasma can offer the flexibility of a simpler 
experimental configuration and less stringent spatio-temporal synchronisation requirements as compared to 
other controlled injection techniques based on secondary laser pulses in orthogonal\cite{Umstadter1996PRL} and
counterpropagating\cite{Faure2006Nature} geometries or with external magnetic-fields\cite{Vieira2011PRL}. 
\nolinebreak[4] Recently electrons injected in the density gradient at the exit of a gas-
jet\cite{Hemker2002PRSTAB,Geddes2008PRL} have been post-accelerated in a capillary-discharge based secondary 
accelerating stage\cite{Gonsalves2011Nature}. The original proposal for density-gradient injection was based 
on density scale lengths greater than the plasma wavelength\cite{Bulanov1998PRE}. Steep gradients in density, 
with scale lengths shorter than the plasma wavelength, can also lead to electron 
injection\cite{Suk2001PRL,Suk2004JOSAB}. Such sharp density gradients have been experimentally generated by 
shock-fronts created with a knife-edge obstructing the flow from the gas-jet nozzle and used to improve the 
quality of accelerated electrons\cite{Koyama2009NIMA,Schmid2010PRST}. 
\paragraph*{} Recent experimental studies\cite{Faure2010PoP} validated the use of plasma perturbation in the 
form of a density-depleted channel for injecting electrons\cite{Hafz2003IEEE,Kim2004PRE} into the wakefield of 
a pump beam propagating across the channel walls. In this article, we report on results extending that 
experiment by changing the position of the plasma channel along the laser wakefield axis. The density depleted
channel was created with a machining laser beam that propagates orthogonal to the pump beam. By changing the
plasma channel position, the injection location was varied and thereby the subsequent accelerating distance. 
This method allowed for inducing controlled electron injection and generating quasi-monoenergetic electron 
beams with a fine control of their energy. The variation of electron energy with acceleration length in our
modified configuration compared to previous experiments\cite{Hsieh2006PRL}, was similarly used to estimate the 
accelerating field strength. It was observed that a threshold plasma length prior to depletion region was 
necessary for density-gradient injection to be effective whereas the final electron beam parameters such as 
energy-spread, divergence and charge were independent of injection location.

%\vspace{-0.4cm}
\section{EXPERIMENTAL SETUP}

\begin{figure}[b]
\includegraphics[width=8.5cm]{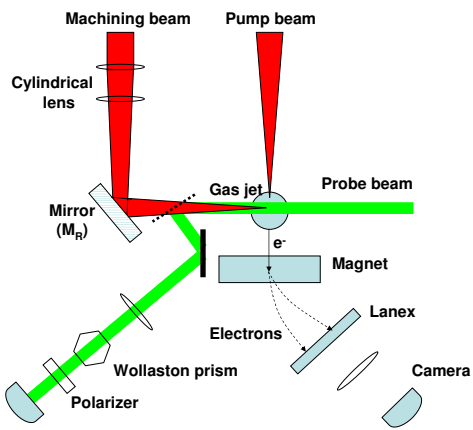}
\caption{\small Schematic figure of the experimental setup : The machining beam, which temporally precedes the 
pump and propagates orthogonally to the pump beam axis, is focused by a pair of cylindrical lens onto the gas-
jet.  A probe beam whose axis is angled to machining beam propagation direction, is used for Wollaston-prism 
based plasma interferometry. Spatial location of the machining focus with respect to the pump beam axis can be 
tuned by the motorised mirror (M$_{\tiny{\textrm{R}}})$. Accelerated electrons from the gas-jet are dispersed 
by the magnet and detected by a Lanex phosphor screen.}
\label{fig:schematic:setup}
\end{figure}

\paragraph*{} The experiments were performed at the Laboratoire d'Optique Appliqu\'ee with the $30$ TW, $30$ 
fs, $10$ Hz, $0.82$ $\mu$m, Ti:Sapphire ``Salle-Jaune" laser system\cite{Pittman2002AppPhyB}. The pump and 
machining beams, propagating orthogonal to each other, are focused onto a supersonic Helium gas-jet ejected 
from a $3$ mm diameter conic nozzle\cite{Semushin2001RSI}. The density profile as characterised by Michelson 
interferometry has a plateau of length $2.1$ mm with $700$ $\mu$m density gradients at the 
edges\cite{Faure2010PoP}. The pump beam with an energy of approximately $0.9$ J is focused by a $70$ cm focal 
length spherical mirror (f$\#\approx 12$) and the machining beam with an energy of $100$ mJ is focused by a 
cylindrical lens system, with a tunable tim{e-d}elay between the two beams. The FWHM spot size of the pump 
beam was estimated to be $14$ $\mu$m $\times$ $18$ $\mu$m with a peak-intensity of approximately $4.3$ 
$\times$ $10^{18}$ W/cm$^2$(normalised vector potential $a_{0}$~$\approx$ $1.5$). The cylindrical focusing 
system for the machining beam consists of two cylindrical lenses (focal lengths of $500$ mm and $400$ mm), 
placed in series such that it generates a line focus of tunable length (FWHM $\sim$ $100$-$200$~$\mu$m) by 
varying the separation between the two lenses. The line focus of the machining beam is oriented in a direction 
orthogonal to the  plane defined by the pump and machining beam axis, whereas the transverse width (FWHM spot-
~size $\approx$ $30$ $\mu$m) of the line is aligned along the pump pulse propagation direction. The peak-
intensity in the line focus estimated to be around $3.4$~$\times$~$10^{16}$~W/cm$^2$ ($a_{0}$~$\approx$ 
$0.1$). The schematic experimental setup consisting of the machining and pump beam along with a probe beam 
(picked off from the pump beam) for Nomarski plasma interferometry\cite{Benattar1979RSI} is shown in Fig. 
\ref{fig:schematic:setup}.

\begin{figure}[b]
\includegraphics[width=8.5cm]{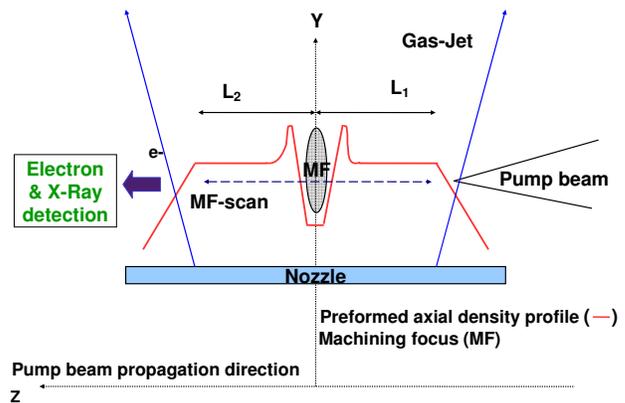}
\caption{\small Schematic figure of the experimental target configuration : Line focus (MF) of the machining 
beam temporally delayed and propagating perpendicular to the pump beam, creates a preformed density-depleted 
channel in the gas-jet. Axial density gradients along the channel walls induce controlled injection of 
electrons into the wakefield of the pump beam. The position of density-depleted channel and thereby the length 
of the plasma interaction region before ($L_{1}$) and after ($L_{2}$) the density depleted zone was varied by 
laterally scanning the machining focus (MF-scan) along the pump beam axis.}\label{fig:schematic:tgtconfig}
\end{figure}

\paragraph*{} The machining laser pulse ionizes the gas-jet and creates a hot plasma localized in the line
focal volume that hydrodynamically expands into the surrounding neutral gas. This leads to the formation of a
density depleted channel with an inner lower density region surrounded by an expanding higher plasma density
channel wall\cite{Milchberg1993PRL,Volfbeyn1999PoP}. Therefore the pump beam sees a longitudinal density-
gradient at the edges of the channel as it propagates in a direction perpendicular to the machining beam. The 
density depletion at the focus of the machining beam creates the axial density gradient (for the pump beam)  
that induces injection of electrons into the wakefield generated by pump pulse. A schematic picture of the 
experimental target configuration with the preformed density-depletion region is shown in 
Fig.~\ref{fig:schematic:tgtconfig}. 
The position of the density depleted channel and thereby the length ($\L_{2}$) of the interaction distance 
following the injection position was tuned by the laterally scanning the machining focus (along the pump beam 
axis) with a motorised mirror (M$_{\tiny{\textrm{R}}}$) placed after cylindrical lens system and 
before the final focus. The aspect ratio of the line focus with an approximate length of $200$ $\mu$m (FWHM) in the vertical direction 
($Y$-axis) and a spot size of $30$ $\mu$m (FWHM) along the pump pulse propagation direction ensures that the 
strongest density gradient as seen by the plasma wakefield of the pump pulse is predominantly longitudinal 
($Z$~-~axis). 
 \nolinebreak[4]

\begin{figure}[b]
\includegraphics[width=8.5cm]{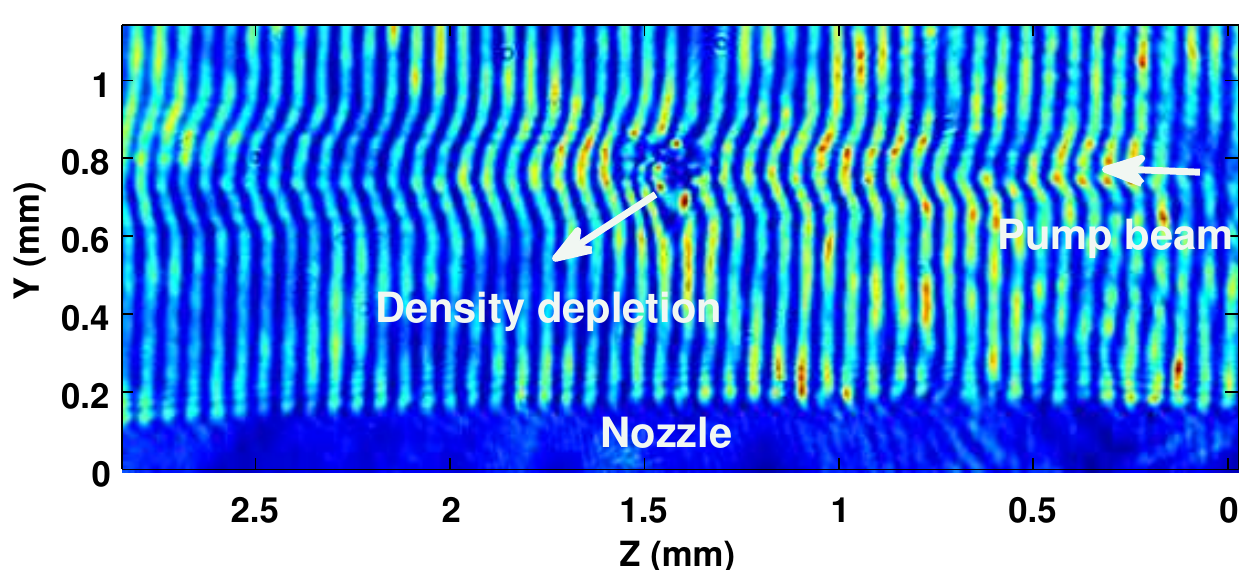}
\caption{\small Interferometric image of the gas-jet : Phase shift due to the plasma generated by the pump 
pulse (propagating right to left in the figure) leads to curved fringes. The density-depleted zone created by 
the machining beam propagating perpendicular (into the plane of figure) to the pump beam gives rise to the 
distortion in the fringes visible in the central region of the interferogram.}
\label{fig:interferogram}
\end{figure}

\paragraph*{} In our experiments, the time delay between the pump and the machining laser pulse was fixed at 
$2$~ns following earlier experiments\cite{Faure2010PoP} where the timing had been optimized to obtain the 
strongest density gradients. Nomarski interferometry allowed us to measure precisely ($\pm$~$50$ $\mu$m) the 
axial location of the density depletion from the position of the distortion in the interferogram fringes (Fig. 
\ref{fig:interferogram}) arising due to the presence of the density channel in the path of the probe beam. 
\nolinebreak
The axial width of the depletion zone estimated from the dimensions of the distorted region was approximately 
$100$-$200$ $\mu$m. However it was not possible to retrieve the longitudinal density profile in the distorted 
region from the interferometers due to the large phase shift caused by the machined plasma. The plasma 
wavelength in our experiments is estimated to be approximately $12$-$15$~$\mu$m corresponding to densities of 
$5-8\times$~$10^{18}$ cm$^{-3}$. Since our conditions are similar to that in Ref. $18$, the density-gradient 
scale length is likewise expected to be similar ($\sim 30~\mu$m) and therefore the change in longitudinal 
density can be considered as gradual compared to plasma wavelength.
As is evident from the integrity of the curved fringes throughout the gas-jet in the left side of the 
interferogram (Z~$>$~$1.4$~mm), the density-depletion or injection zone in the path of the pump beam does not 
appear to disrupt its subsequent propagation and self-guiding. The pump pulse appears to be guided for lengths 
greater than $980\mu$m (Rayleigh range for a Gaussian focal spot size of $16$~$\mu$m) both before and after 
the depletion zone. Controlled injection of electrons into the plasma wakefield occurs in the density-
gradients of the density-depleted region and acceleration in the subsequent homogeneous plasma.

\paragraph*{} The energy of the accelerated electron bunch exiting the gas-jet was measured with a magnetic 
spectrometer consisting of $1.1$ Tesla, $10$ cm magnet and a Lanex phosphor screen imaged onto a $16$-bit CCD 
camera. The spectrometer energy resolution was $2.3\%$ at $100$ MeV. The electron energy spectrum and absolute 
charge is obtained by post-processing the recorded data taking into account the calibration of the 
diagnostic\cite{Glinec2006RSI}.

\section{RESULTS \& DISCUSSION}

\begin{figure}[b]
\centering
\includegraphics[width=9.5cm,height=1.6cm]{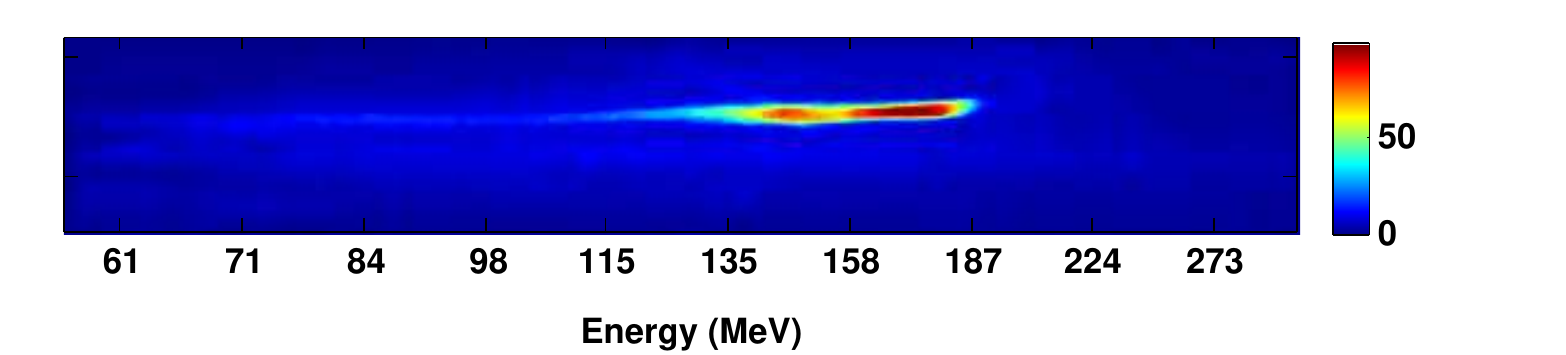}
\includegraphics[width=8.5cm,height=5.5cm]{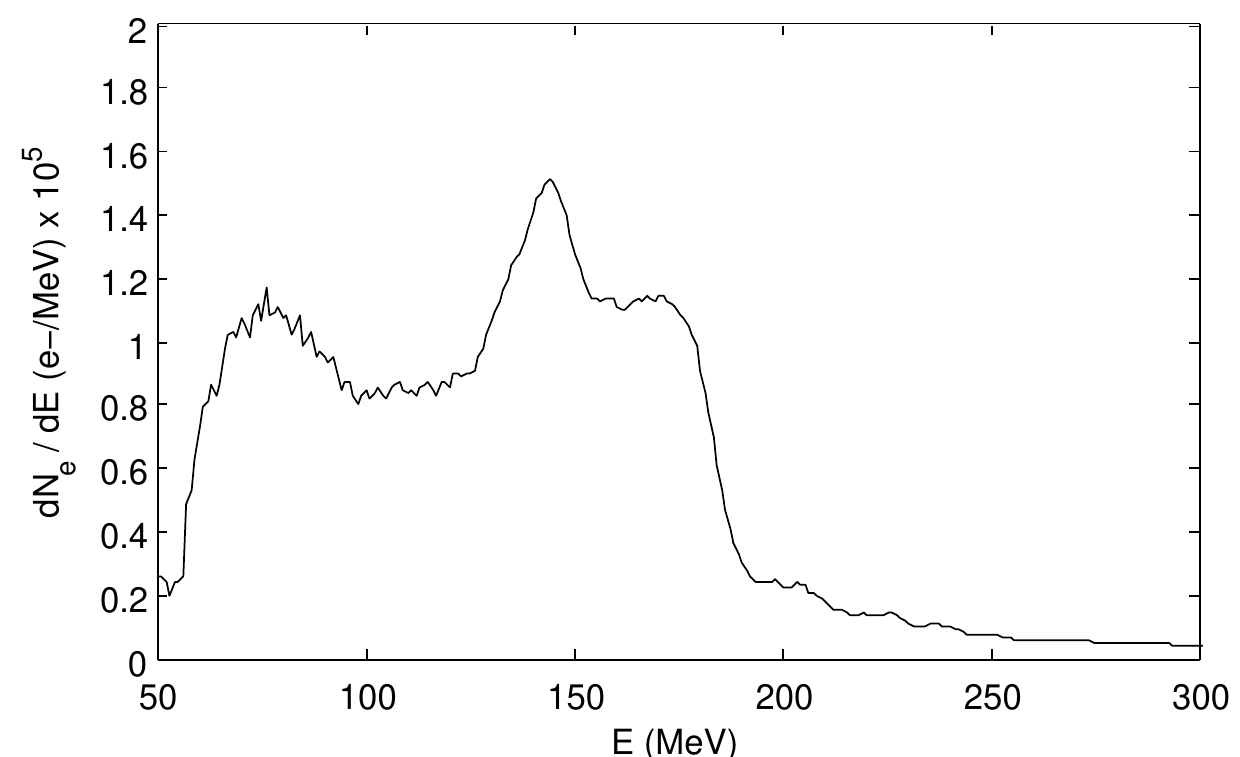}
\caption{\small Raw image of the electron beam on the Lanex screen (top) and lineout of the corresponding 
electron spectrum (bottom) obtained from a homogeneous plasma with only the pump beam. Electrons are generated 
by self-injection with a large energy spread for a plasma density of approximately $8$ $\times$ ${10}^{18}$
cm$^{-3}$.}
\label{fig:filename_pumponlyspectra}
\end{figure}

\paragraph*{} In our experiments, electrons are trapped and accelerated by the strong electric fields of the 
nonlinear plasma wave (plasma bubble)\cite{Pukhov2002AppPhyB} excited by the intense and ultrashort pump laser 
pulse propagating in the gas-jet. Depending on specific plasma conditions, injection into the bubble can occur 
either by self-trapping or by controlled injection due to the preformed density-perturbation. The acceleration 
of electrons occurs in the matched plasma bubble regime of resonant laser wakefield acceleration 
\cite{Malka2002Science,Mangles2004Nature,Faure2004Nature}(pulse length $\leq$ plasma wavelength/2) wherein the 
laser focal spot size is comparable to the bubble radius which is approximately of the order of half a plasma 
wavelength\cite{Lu2007PRSTAB}. In order to differentiate between self-injection and density-gradient 
injection, electron spectrum was first recorded with only the pump beam focused onto the gas-jet. The 
density was reduced to minimize self-injection as much as possible without complete reduction of the detected 
charge. In these conditions (electron densities of about $5$~-~$8$~$\times$~${10}^{18}$~cm$^{-3}$), self-
injection occurs occasionally, resulting in the production of a poor quality electron beam with a broadband 
electron energy distribution. The Lanex image and the corresponding electron spectrum for one such target shot 
are shown in Fig. \ref{fig:filename_pumponlyspectra}. The spectrum of the self-injected electrons is 
consistently characterized by a large energy spread, low-energy dark current and considerable fluctuations  in 
the spectral profile with high backround level on consecutive shots. The mean value of maximum electron energy 
(defined as cut-off edge in  the logarithm of the spectral profile) was around $170-175$~MeV.

\begin{figure}[t]
\includegraphics[width=8.5cm]{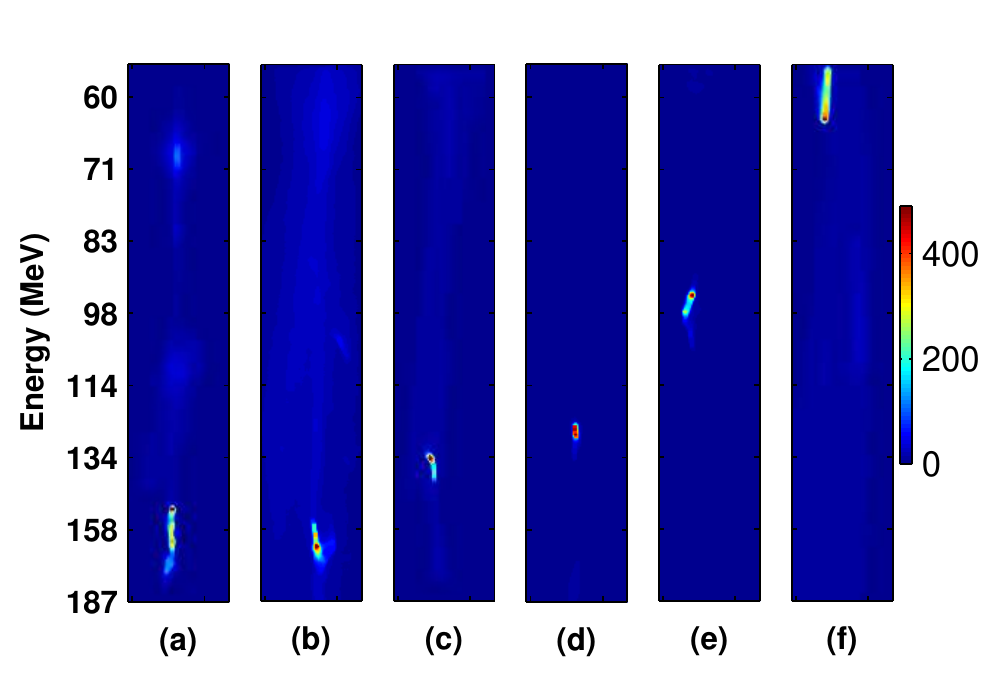}
\caption{\small Raw images of the electron beam on the Lanex screen obtained by injection at different axial 
locations ($Z_{m}$) of the machining focus.  $Z_{m}$ = (a) $1.6$ mm (b) $1.8$ mm (c) $1.9$ mm (d) $2$ mm (e)
$2.2$ mm (f) $2.5$ mm from the entrance of the gas-jet.}
\label{fig:spectrastack}
\end{figure}

\begin{figure}[t]
\includegraphics[width=8.5cm]{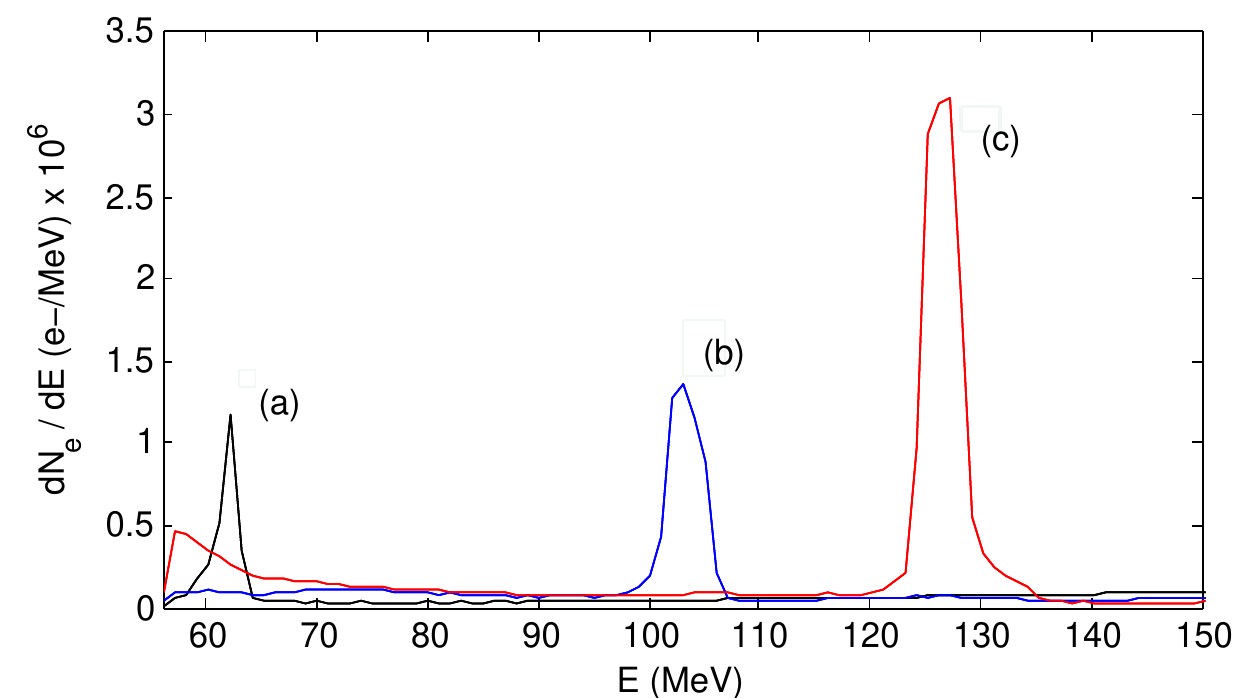}
\caption{\small Experimental quasi-monoenergetic electron spectrum with relative energy spread 
($\Delta{E_{\small{{{F}{W}{H}{M}}}}}/E_{peak}$)  of around $3\%$ obtained by densit{y-g}radient injection for 
three different axial location (Z$_{m}$) of the machining focus. Z$_{m}=$ (a) $2.5$ mm (b) $2.2$ mm (c) $2.0$ 
mm from the entrance of the gas-jet. The injected charge is (a) $0.2$ pC   (b) $1$ pC  (c) $2.2$ pC and the 
plasma density is approximately $8$ $\times$ ${10}^{18}$ cm$^{-3}$.}
\label{fig:bsdspectrum}
\end{figure}

\noindent The electron signal in the high-energy tail ($> 175$ MeV) of the spectrum 
(Fig.~\ref{fig:filename_pumponlyspectra}) is due to the high background level. For plasma densities much below 
the self-injection threshold, there were no distinct electron peaks with significant charge and the detected 
electron distribution was very close to the background level. The presence or absence of density-depleted 
region under these conditions did not have any significant effect on the electron spectrum because the 
densities are too low to excite a wakefield of sufficient amplitude that can trap and accelerate electrons. 
Self-injection at low densities would require laser system with greater power. At higher densities 
($\sim$~$5$~-~$8$~$\times$~$10^{18}$~cm$^{-3}$), there is an increased probability of 
intermittent self~-~trapping of electrons with a poor accelerated beam quality. However for the same 
experimental conditions, firing the machining beam resulted in significant improvement of electron spectrum. 
The effect of the depletion region, in the form of localized electron injection at the density-gradients, 
dominates over any occasional self~-~injection. In contrast to the case of self~-~injection in a homogeneous 
plasma, the presence of the preformed density perturbation, leads to the acceleration of electrons with low 
energy spread, indicating the benefits of controlled injection for a fixed laser power. 
\paragraph*{} The axial location of the depletion region was varied by translating the position of the 
machining focus from the entrance to the exit of the gas-jet along the direction of the pump laser-axis. 
Electron spectrum data was recorded by scanning the location of the depletion region in steps of approximately 
$0.1$ mm and keeping all other experimental conditions unchanged. Electron beam images on the Lanex screen, 
obtained on selected shots for the scanned axial locations ($Z_{m}$) of the machining focus in the range of 
$1.6$ mm to $2.5$ mm, are shown in Fig.~\ref{fig:spectrastack}. \nolinebreak[4]
\noindent By changing the location of the density depletion and thereby the subsequent plasma interaction 
length, the final energy of the accelerated electrons is observed to be tunable. The spectra at different 
longitudinal locations ($Z_{m}$) of the machining focus are shown in Fig.\hspace{3pt}\ref{fig:bsdspectrum}. 
The relative energy spread ($\Delta{E_{\small{{{F}{W}{H}{M}}}}}/E_{peak}$) of the quasi-monoenergetic peaks is 
around $3\%$, limited by electron spectrometer in these few selected shots. Moreover the peak signal level in the density-gradient injected spectrum is approximately ten times higher than in the case of the self-injected electron spectrum. The 
improvement in electron beam quality with the machining beam highlights the advantages of controlled injection 
over uncontrolled self-injection, besides offering the flexibilty of tuning the electron energy with a single 
gas-jet in this particular experimental geometry. 

\begin{figure}[t]
\includegraphics[width=8.5cm,height=5.5cm]{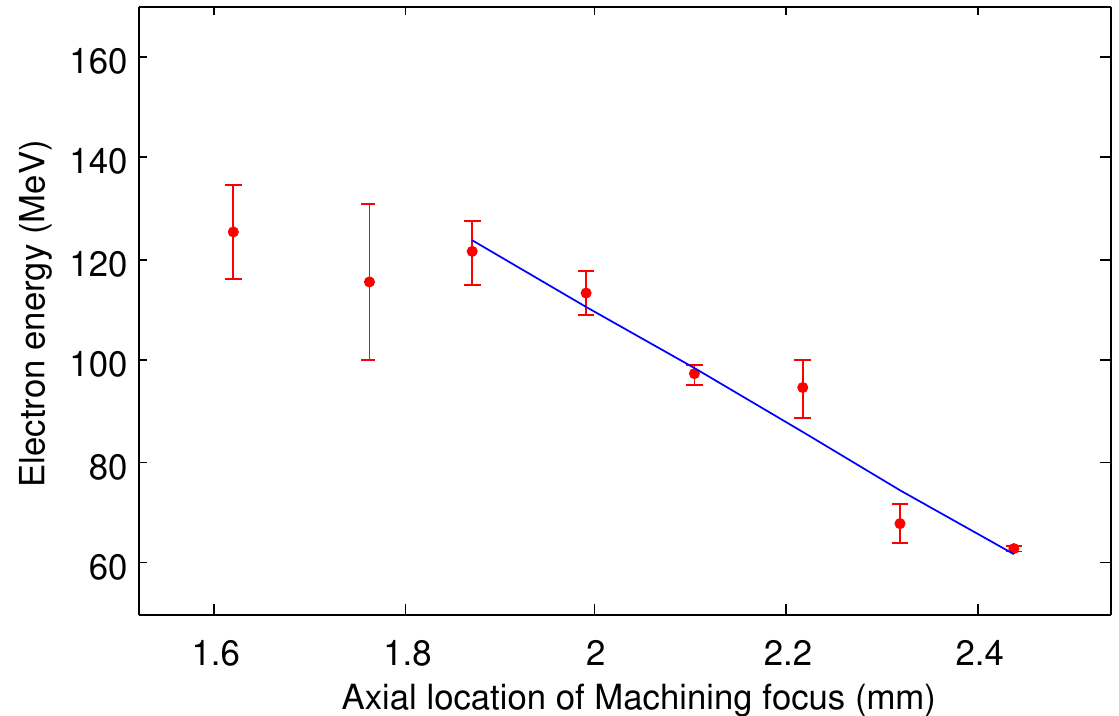}
\caption{\small{}Variation of the mean energy of the quasi-monoenergetic electrons with the axial location of 
the machining focus (density-depleted zone) that induces density-gradient injection. Energy was tunable from a 
maximum of $120$ MeV to $60$ MeV by varying the location of the machining focus in the range of $1.9$ mm to 
$2.45$ mm from the entrance of the gas-jet. Dots are the mean of accumulated data from multiple shots and 
error bars are the standard error of the mean.}
\label{fig:energylength}
\end{figure}

\paragraph*{} Quasi-monoenergetic electrons were not detected when the depletion region was placed closer to 
the entrance in the first-half of the gas-jet, indicating that there is a threshold pump-pulse propagation 
distance after which the electrons begin to get injected in the longitudinal gradients of the depletion zone.
The threshold length for our experimental conditions in the case of density-gradient injection was found to be 
approximately $1.4$ mm from the entrance of the gas-jet. Initially the focussed pump laser 
pulse has intensity ($a_{0} \sim 1.5$) and parameters (spot size $\sim$~$16$~$\mu$m, pulse length 
$\sim$~$9$~$\mu$m) that are far from the matched, plasma bubble regime of resonant laser wakefield 
acceleration. For the matched regime at our plasma densities, the pump spot size has to be approximately equal 
to the bubble radius ($\simeq$ $6-10$~$\mu$m). Therefore a long interaction distance is needed for the laser 
pulse to be sufficiently compressed transversally and longitudinally in order to drive a suitable nonlinear 
plasma wave for trapping electrons. Through an interplay of self-focusing, pulse-shortening and self-
steepening, the spot size and the temporal duration of the pump laser pulse evolves to reach the matched 
regime after propagating for a certain axial distance from the focal 
position\cite{Malka2002Science,Faure2005PRL,AGR2007PRL}. Quasi-static WAKE simulations\cite{Mora1997PoP} 
reveal that for a pump laser focus location in the range of [$0-700$] $\mu$m on the edge of the gas-jet, the 
initial normalized laser amplitude ($a_{0}$) of $1.5$ increases to a maximum value ($a_{l}$) of about 
$3.2-3.6$ after a propagation distance of approximately $1+$/$-0.1$ mm, close to the experimentally observed 
threshold length. The increased laser amplitude is due to initial focal spot size and pulse duration 
compressing to approximately $8~\mu$m and $22$~fs respectively. Though the exact values of the final laser 
parameters can change with the initial focus location, they are approximately close to the matched regime and 
favours the generation of a plasma bubble that can trap electrons as it traverses the density-depleted region. 
When the depletion region was placed closer to the exit of the gas-jet, quasi-monoenergetic electrons were 
observed with lower peak energy compared to the case when focused at the center. For propagation lengths 
($L_{1}$) in the range of $1.4$ mm to $1.8$ mm, the spectrum had greater instability compared to case of 
lengths greater than $1.8$ mm presumably due to conditions being close to the matched regime or thresholds of 
electron injection. For the data set obtained with the machining beam, the probability of injection for which 
a quasi-monoenergetic electron distribution has been measured was approximately $50\%$. In the other $50\%$ of 
case, no electrons were observed or when they were measured exhibited a broad energy distribution with a lower 
total charge. The shot to shot stability could be improved in the future by tuning the delay between the pump 
and the machining pulse (or the machining laser energy). The probability of injection could also be improved 
by better control over laser conditions that were perhaps sub-optimal during this particular experiment. \nolinebreak[4]
%\vspace{+0.5cm}
\noindent \paragraph*{} In Fig. \ref{fig:energylength}, the final electron energy is plotted as a function of 
the density-depletion position $L_{1}$. The data points are the mean of accumulated data from multiple shots 
and the straight line is a fit over the data corresponding to the $L_{1}$ in the range $1.9$ mm to $2.45$ mm. 
Since the axial location of machining focus ($L_{1}$) determines the plasma interaction length ($L_{2}$) 
following the depletion region (see Fig. \ref{fig:schematic:tgtconfig}), the maximum possible acceleration 
length ($L_{acc}$) in a $3$ mm gas-jet approximately equals $L_{2} \approx 3-L_{1}$ mm. Note that for our 
experimental conditions, the net acceleration length is less than the maximum possible value since the 
density-gradient injection is effective only for $L_{1} \geq $ $1.4$ mm. The linear region in the graph (Fig. 
\ref{fig:energylength}) quantifies the tunability of electron energy with acceleration length. The final 
electron energy on an average varied from $120$ MeV to $60$ MeV for an acceleration length varying from 
$1.2$~mm to $0.6$~mm equivalent to an acceleration gradient of $100$~GeV/m. This value is similar to that 
measured recently in colliding pulse injection\cite{Corde2011PRL} but lower than that expected from 
theory\cite{Lu2007PRSTAB}. The scaling law predicts an acceleration gradient ($\sim 48{a_{l}}^{1/2}
{n_{e}}^{1/2}$) of approximately $192-256$ GeV/m for our parameters, which is greater than the experimental 
measurement by a factor of $2-2.5$. This could be probably due to a decrease of the laser intensity in the 
second half of the gas jet or the deformation of the bubble resulting from laser pulse evolution that reduces 
the electron energy gain. The acceleration length ($L_{acc}$) in our case is limited to about $1$ mm. For 
machining focus locations ($L_{1}$) prior to $1.9$ mm, there appears to be a trends towards saturation of the 
mean electron energy. In this region, the spectrum is unstable with larger fluctuations in peak energy 
compared to the linear region. In some shots multiple peaked spectrum with energies as high as $140-170$ MeV 
in the highest energy peak, whereas in certain shots single peaks with much lower energies were observed. This 
could be due to $a_{o}$ evolution during laser propagation and the possibility of occurrence of multiple bunch 
injection on the densit{y-g}radient. The energy spectrum data was also analysed by plotting the maximum cut-
off energy and a similar trend was observed with a slightly greater slope in the linear region. 

\begin{figure}[t]
\includegraphics[width=8.5cm,height=6.0cm]{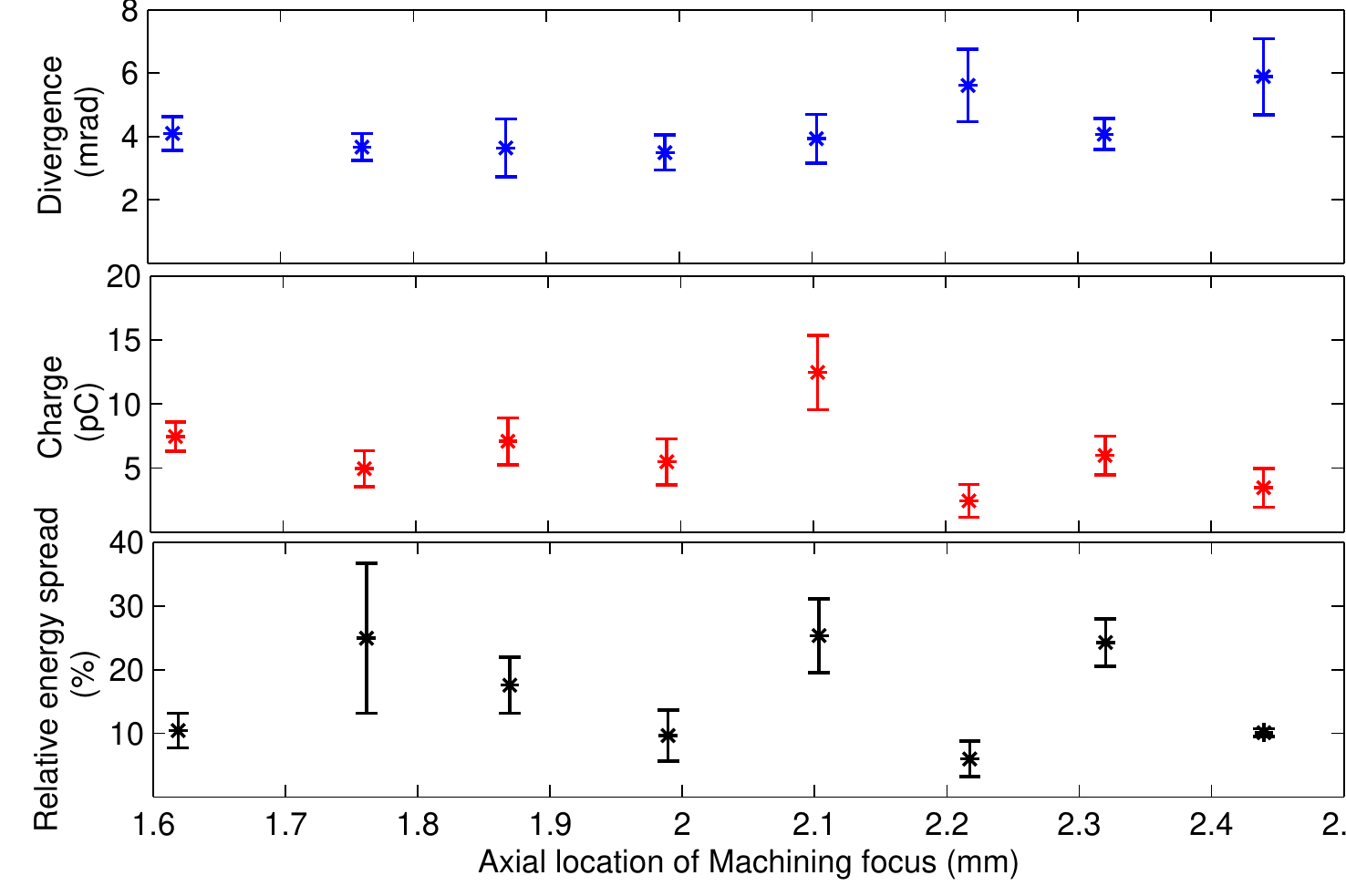}
\caption{\small Variation of the divergence, charge and relative energy spread of the quasi-monoenergetic 
electrons with the axial location of the machining focus. The divergence, injected charge and relative energy 
spread is relatively constant with approximate mean values ($\pm$ std.~error) of $4 \pm 0.8$ mrad, $6 \pm 1.8$ 
pC and $15 \pm 3.6\%$ respectively. Dots are the mean of accumulated data from multiple shots and error bars 
are the standard error of the mean.}
\label{fig:parameters}
\end{figure}

\paragraph*{} Finally the variation of various electron beam parameters such as relative energy spread, 
divergence and charge within the peaks of the spectrum were analysed as a function of the axial location of 
the machining focus (Fig. \ref{fig:parameters}). The divergence, injected charge and relative energy spread is 
relatively constant across the acceleration length with approximate mean values ($\pm$ std.~error) of $4 \pm 
0.8$ mrad, $6 \pm 1.8$ pC and $15 \pm 3.6\%$ respectively.

\section{CONCLUSION}

In summary, quasi-mononenergetic electrons were generated by using the longitudinal density gradients of a 
plasma density perturbation to induce controlled injection of electrons into the plasma wakefield. The density 
perturbation in the form of a density depleted plasma channel was created by a secondary machining beam. Final 
energy of the accelerated electrons was tuned from a maximum of $120$ MeV to $60$ MeV by varying the axial 
position of the density perturbation and thereby the injection location and the subsequent plasma interaction 
length. A threshold plasma length prior to depletion region was required for density-gradient injection to be 
effective whereas the final electron beam parameters such as energy-spread, divergence and charge were 
observed to be independent of the injection location. Controlled injection in a longitudinally inhomogeneous 
plasma appears better than sel{f-i}njection in a homogeneous plasma in terms of final electron-beam quality 
for the same experimental conditions and offers the flexibility of tuning the electron energy with a single 
gas-jet. In future, accurate measurements of the density profile in the density-gradient injection scheme will 
allow for benchmarking numerical simulations to optimize the experimental parameters needed for generating 
high-quality electron beams.

\section*{ACKNOWLEDGMENTS}

We thank J.P. Goddet and A. Tafzi for the operation of the laser system. We acknowledge the support of the 
European Research Council for funding the PARIS ERC project (Contract No. 226424), EC FP7 LASERLAB-
EUROPE/LAPTECH (Contract No. 228334) and EU Access to Research Infrastructures Programme Project LASERLAB-EUROPE II.

\end{document}